\documentclass[twocolumn]{aastex62}

\submitjournal{ApJ}

\shorttitle{Radial and vertical distributions of DCN and DCO$^+$}
\shortauthors{\"Oberg et al.}

\begin{document}

\title{The TW Hya Rosetta Stone Project I: Radial and vertical distributions of DCN and DCO$^+$}

\correspondingauthor{Karin \"Oberg}
\email{koberg@cfa.harvard.edu}

\author{Karin I. \"Oberg}
\affil{Harvard-Smithsonian Center for Astrophysics, 60 Garden St., Cambridge, MA 02138, USA}

\author{L. Ilsedore Cleeves}
\affil{Department of Astronomy, University of Virginia, Charlottesville, VA 22904, USA}

\author{Jennifer B. Bergner}
\affil{NASA Sagan Fellow, University of Chicago Department of the Geophysical Sciences, Chicago, IL 60637, USA}

\author{Joseph Cavanaro}
\affil{Harvard-Smithsonian Center for Astrophysics, 60 Garden St., Cambridge, MA 02138, USA}

\author{Richard Teague}
\affil{Harvard-Smithsonian Center for Astrophysics, 60 Garden St., Cambridge, MA 02138, USA}

\author{Jane Huang}
\affil{Harvard-Smithsonian Center for Astrophysics, 60 Garden St., Cambridge, MA 02138, USA}
\affil{NHFP Sagan Fellow, Department of Astronomy, University of Michigan, 323 West Hall, 1085 S. University Avenue, Ann Arbor, MI 48109, USA}

\author{Ryan A. Loomis}
\affil{National Radio Astronomy Observatory, Charlottesville, VA 22903, USA}

\author{Edwin A. Bergin}
\affil{Department of Astronomy, University of Michigan, 1085 S. University Ave, Ann Arbor, MI 48109}

\author{Geoffrey A. Blake}
\affil{Division of Chemistry \& Chemical Engineering, California Institute of Technology, Pasadena CA 91125}

\author{Jenny Calahan}
\affil{Department of Astronomy, University of Michigan, 1085 S. University Ave, Ann Arbor, MI 48109}

\author{Paolo Cazzoletti}
\affil{Leiden Observatory, Leiden University, PO Box 9513, 2300 RA Leiden, The Netherlands}

\author{Viviana Veloso Guzm\'an}
\affil{Instituto de Astrof\'isica, Ponticia Universidad Cat\'olica de Chile, Av. Vicu\~na Mackenna 4860, 7820436 Macul, Santiago, Chile}

\author[0000-0001-5217-537X]{Michiel R. Hogerheijde}
\affiliation{Leiden Observatory, Leiden University, PO Box 9513, 2300 RA Leiden, The Netherlands}
\affiliation{Anton Pannekoek Institute for Astronomy, University of Amsterdam, Science Park 904, 1098 XH, Amsterdam, The Netherlands}

\author{Mihkel Kama}
\affil{Institute of Astronomy, University of Cambridge, Madingley Road, Cambridge CB3 0HA, UK}

\author[0000-0002-3800-9639]{Jeroen Terwisscha van Scheltinga}
\affiliation{Laboratory for Astrophysics, Leiden Observatory, Leiden University, PO Box 9513, 2300 RA Leiden, The Netherlands}
\affiliation{Leiden Observatory, Leiden University, PO Box 9513, 2300 RA Leiden, The Netherlands}

\author{Chunhua Qi}
\affil{Harvard-Smithsonian Center for Astrophysics, 60 Garden St., Cambridge, MA 02138, USA}

\author{Ewine van Dishoeck}
\affil{Leiden Observatory, Leiden University, PO Box 9513, 2300 RA Leiden, The Netherlands}

\author{Catherine Walsh}
\affil{School of Physics and Astronomy, University of Leeds, Leeds LS2 9JT, UK}

\author{David J. Wilner}
\affil{Harvard-Smithsonian Center for Astrophysics, 60 Garden St., Cambridge, MA 02138, USA}

\begin{abstract}

Molecular D/H ratios are frequently used to probe the chemical past of Solar System volatiles. Yet it is unclear which parts of the Solar Nebula hosted an active deuterium fractionation chemistry. To address this question, we present 0\farcs2--0\farcs4 ALMA observations of DCO$^+$ and DCN 2--1, 3--2 and 4--3 towards the nearby protoplanetary disk around TW Hya, taken as part of the TW Hya Rosetta Stone project, augmented with archival data. DCO$^+$ is characterized by an excitation temperature of $\sim$40~K across the 70~au radius pebble disk, indicative of emission from a warm, elevated molecular layer. Tentatively, DCN is present at even higher temperatures. Both DCO$^+$ and DCN present substantial emission cavities in the inner disk, while in the outer disk the DCO$^+$ and DCN morphologies diverge: most DCN emission originates from a narrow ring peaking around 30~au, with some additional diffuse DCN emission present at larger radii, while DCO$^+$ is present in a broad structured ring that extends past the pebble disk. Based on parametric disk abundance models, these emission patterns can be explained by a near-constant DCN abundance exterior to the cavity, and an increasing DCO$^+$ abundance with radius. There appears to be an active deuterium fractionation chemistry in multiple disk regions around TW Hya, but not in the cold planetesimal-forming midplane and in the inner disk. More observations are needed to explore whether deuterium fractionation is actually absent in these latter regions, and if its absence is a common feature, or something peculiar to the old TW Hya disk.

\end{abstract}


\section{Introduction  \label{sec:intro}}

Solar System volatiles and organics are often observed to have non-Solar isotopic compositions, especially with respect to their D/H ratios \citep{Robert82,Mumma11,Ceccarelli14,Alexander17,Altwegg19}. The observed deuterium enrichments encode information about the molecules' formation environment. They can therefore be used to differentiate between inheritance of material from the pre-Solar molecular cloud, and {\it in situ} formation in the Solar Nebula \citep[e.g.][]{Cleeves14}. Differences in D/H ratios between different Solar System reservoirs have also been used to constrain the origins of Earth's water \citep{Hartogh11,Altwegg15}, and to infer that Solar Nebula chemistry, especially in the inner Solar Nebula, resulted in a reduction of D/H levels in inherited, initially deuterium-rich volatiles  \citep{Yang13,Furuya17}. Despite decades of Solar System measurements and models, our understanding of deuterium fractionation chemistry, and the distribution of deuterated volatiles in the Solar Nebula are, however, limited. Observations of deuterated species in analogs to the Solar Nebula, i.e., in protoplanetary disks, are key to anchor our models of deuterium fractionation chemistry in disks, and to put Solar System measurements in context.

To date, four deuterated isotopologues have been detected in protoplanetary disks at millimeter wavelengths, DCO$^+$, DCN, N$_2$D$^+$, and C$_2$D \citep{Dutrey96,vanDishoeck03,Thi04,Guilloteau06,Qi08,Fuente10,Oberg10c,Oberg11a,Oberg12,Mathews13,Teague15,Huang15,Qi15,Oberg15b,Huang17,Salinas17,Loomis20}. When observed, the distributions of deuterated and non-deuterated isotopologues are frequently different, which implies some {\it in situ} formation \citep{Huang17}. In other words, these observations show that there is an active deuterium fractionation chemistry in at least some disk locations during planet formation. Whether this chemistry impacts the deuterium levels in volatiles in the disk midplane where planetesimals assemble depends on which disk layer is producing the observed deuterated isotopologue emission, and the efficiency of vertical mixing in disks.

{\it A priori} DCO$^+$ and DCN emission may originate from a range of disk layers because there are multiple potential formation pathways for each molecule \citep[e.g.][]{Millar91,Aikawa99,Turner01,Willacy07,Favre15,Aikawa18,Roueff15}.  In general, pathways that begin with the formation of H$_2$D$^+$ are active at low temperatures, $<30$~K, characteristic of regions close to the disk midplane, while pathways that begin with deuterated small hydrocarbons, initiated by the CH$_2$D$^+$ ion, can operate at a larger range of temperatures \citep{Wootten87,Parise09,Roueff15}, including in inner disk regions and in disk atmospheres. Most spatially resolved observations of DCO$^+$ show extended emission in the outer disk, and a lack of emission in the inner disk \citep{Qi08,Oberg15b,Huang17,Salinas17}, which is most consistent with formation through the colder H$_2$D$^+$ pathway. This conclusion was supported by a  measurement of the DCO$^+$ excitation temperature in the disk around HD 163296, which was estimated to 12-20~K \citep{Flaherty17}. However, contributions from the CH$_2$D$^+$ pathway cannot generally be excluded, and \citet{Carney18} found that in the case of the disk around HD 169142, the majority of observed DCO$^+$ emission originates from a warmer disk region. DCN emission generally, but not always, appears radially more compact than DCO$^+$ \citep{Oberg12,Huang17,Salinas17}, indicative of a larger contribution from the warmer formation pathways. There are no direct measurements of the DCN excitation temperature in a disk.  

In this paper, we use ALMA observations of multiple DCO$^+$ and DCN lines to map out the DCO$^+$ and DCN distributions and excitation temperatures across the nearby protoplanetary disk around TW Hya. TW Hya is an excellent bench-marking system because of extensive structural and chemical modeling \citep[e.g.][]{Bergin13,Cleeves15}.
The present study uses data from the TW Hya Rosetta Stone Project along with other archival date sets. The Rosetta program set out to map chemistry at 10 AU resolution to understand the spatial distribution of commonly observed molecules and their isotopologues towards TW Hya. The project as a whole will inform studies at lower resolution, which in some cases is unavoidable for more distant protoplanetary disks. This paper is organized as follows: \S \ref{sec:obs} summarizes the observational details and the data reduction procedure. Section \ref{sec:results} presents the DCO$^+$ and DCN observational data products, disk-averaged and radially resolved column densities and excitation temperatures using rotational diagrams. Informed by the rotational diagram analysis, \S\ref{sec:models} introduces a series of toy models, generated using RADMC-3D \citep{Dullemond12}, aimed at qualitatively exploring what kinds of 2D abundance structures can explain the observations. We discuss the observations and modeling results in \S \ref{sec:disc}, followed by a summary and some concluding remarks in \S\ref{sec:conc}. 		

\section{Observational Details\label{sec:obs}}

\begin{deluxetable*}{lcccccccc}
\tablecaption{Observational details of DCO$^+$ and DCN lines\label{tab:obs}}
\tablehead{
\colhead{Line}    &\colhead{Date}   &\colhead{Int. Time} & \colhead{\# Ant.}    & \colhead{Baselines}	 & \colhead{Ang. Res.}& \colhead{Max Ang. Scale}& \colhead{Phase Cal.}& \colhead{Flux Cal.}\\
 \colhead{} & \colhead{}    & \colhead{min} &
\colhead{}    &\colhead{m} &\colhead{$\arcsec$}&\colhead{$\arcsec$}& \colhead{}& \colhead{} 
}
\startdata
J=2--1  & Oct 22, 2016 & 46 & 48 & 19--1396 & 0.37 & 3.8 &J1037-2934   &J1037-2934\\
        & Oct 25, 2016 & 46 & 48 & 19--1396 & 0.37 & 3.8 &J1037-2934   &J1107-4449\\
        & Oct 27, 2016 & 46 & 48 & 19--1396 & 0.37 & 3.8 &J1037-2934   &J1107-4449\\\\
        J=3--2  &Dec 16, 2016   &24   & 45 &15--460   & 0.71 &6.4 &J1037-2934 &J1107-4449\\
        &May 5, 2017    &40   &45 &15--1124   & 0.29 &4.2  &J1037-2934   &J1037-2934\\
        &May 7, 2017    &40   &50  &15--1124   & 0.28 &3.9  &J1037-2934   &J1037-2934\\\\
        J=4--3  &Feb 1, 2017   &29   & 41 &15--260   & 0.89 & 7.6&J1037-2934 &J1107-4449\\
        &Jan 23, 2018    &48   & 43 &15--1398   & 0.20 & 2.9 &J1037-2934   &J1037-2934\\
        &Sep 20, 2018    &48   &44  &15--1398   & 0.20 &3.0 &J1037-2934   &J0904-5735\\
\enddata
\end{deluxetable*}

\begin{deluxetable*}{lccccccccc}
\tablecaption{Line catalogue\tablenotemark{a} and observational data\label{tab:line-data}}
\tablehead{
\colhead{Line}   & \colhead{Rest freq.}    & \colhead{Log$\rm_{10}(A_{ij})$	}& \colhead{E$_{\rm u}$}    & \colhead{$g_{\rm u}$	} & \colhead{beam (PA)}	 & \colhead{rms\tablenotemark{b}} & \colhead{Flux\tablenotemark{c}}& \colhead{Flux $<$100 au}& \colhead{$\Delta$v\tablenotemark{f}}\\
 \colhead{}   & \colhead{GHz}    & \colhead{} & \colhead{K}    &\colhead{} & \colhead{$\arcsec\times\arcsec$ ($^{\circ}$)} &\colhead{mJy beam$^{-1}$}& \colhead{mJy km/s}& \colhead{mJy km/s} & \colhead{km/s}
}
\startdata
DCO$^+$ J=2--1  &144.077285\tablenotemark{d}    &-3.67   &10.37  &5&$0\farcs42\times0\farcs34$ ($84^{\circ}$)	&1.2    &559$\pm$16 &471$\pm$11     &2.1--3.5\\
DCO$^+$ J=3--2  &216.112582\tablenotemark{d}    &-3.12   &20.74  &7&$0\farcs49\times0\farcs31$ ($88^{\circ}$)	&3.4    &1904$\pm$19 &1532$\pm$13     &2.1--3.5\\
DCO$^+$ J=4--3  &288.143858\tablenotemark{d}    &-2.73   &34.57  &9&$0\farcs24\times0\farcs22$ ($80^{\circ}$)	&2.4    &3622$\pm$33 &3003$\pm$22     &2.1--3.5\\\\
DCN J=2--1      &144.827996\tablenotemark{d}    &-3.89   &10.42  &15&$0\farcs42\times0\farcs34$ ($85^{\circ}$)	&1.1    &$>$128$\pm$12\tablenotemark{e} &$>$101$\pm$8\tablenotemark{e}      &2.1--3.5\\
DCN J=3--2      &217.238537\tablenotemark{d}    &-3.34   &20.85  &21&$0\farcs48\times0\farcs31$ ($88^{\circ}$)	&3.3    &641$\pm$20 &561$\pm$13     &2.2--4.0\\
DCN J=4--3      &289.644917\tablenotemark{d}    &-2.95   &34.75  &27&$0\farcs24\times0\farcs22$ ($80^{\circ}$)	&2.3    &1344$\pm$29 &1206$\pm$19     &2.0--3.7\\
\enddata
\tablenotetext{a}{Line catalogue data from CDMS \citep{Muller05}}
\tablenotetext{b}{In 0.07 km/s channels for 3--2 and 4--3 lines, and in 0.35 km/s channels for 2--1 lines}
\tablenotetext{c}{Integrated out to 2\farcs5. Uncertainty does not include 10\% flux calibration uncertainty.}
\tablenotetext{d}{There is no visible DCO$^+$ and DCN fine structure in our data.}
\tablenotetext{e}{The DCN 2--1 line emission is likely underestimated due to low SNR and some of the emission being carried by fine structure lines, which are not included.}
\tablenotetext{f}{The velocity range over which emission is integrated.}
\end{deluxetable*}

DCO$^+$ and DCN J=3--2 and 4--3 observations towards TW Hya were acquired as a part of the Rosetta Stone project (2016.1.00311.S and 2017.1.00769.S, PI: Cleeves) in six separate executions. The two J=3-2 lines were observed together in one Science Goal, and the two J=4-3 lines in a second one. Observations of DCO$^+$ and DCN 2--1 were obtained as part of project 2016.1.00440.S (PI: Teague) in three separate executions. The observation dates, integration times, number of antennas, range of baselines, nominal angular resolutions, maximum recoverable angular scales, phase calibrators and flux calibrators are listed in Table \ref{tab:obs}

The measurement sets associated with each execution was initially calibrated by JAO staff. Additional self calibration was applied to the pipeline calibrated data for the TW Hya Rosetta Stone observations using CASA 4.5. Phase-only self calibration was conducted on the line free continuum using 30 second integrations and averaging both polarizations. The solutions were applied to the full resolution measurement sets. 
The line data were continuum subtracted using {\it uvcontsub} and imaged with the CLEAN algorithm \citep{Hogbom74}. The 3--2 and 4--3 line observations were cleaned with 0.07 km/s velocity resolution and the 2--1 with 0.35 km/s resolution down to a level of $2\times$rms. During the CLEANing process, we employed a mask, constructed by manually identifying areas with emission in each channel, and a Briggs parameter of 0.5 for the fiducial image cubes. The resulting beam sizes, and peak line emission and rms in Jy/beam are presented in Table \ref{tab:line-data}. 

We also produced a second set of image cubes with the resolution of all lines smoothed using {\it imsmooth} in CASA with a circular 0\farcs5 beam. This resolution matches the major axis of the beams in the 3--2 line data and thus constitute the highest uniform resolution we can achieve across the sample with a circular beam. These image cubes are used in the quantitative line analysis in \S\ref{sec:rot-dia} where a uniform resolution for the 2--1, 3--2 and 4--3 lines is required. 

\section{Results \label{sec:results}}

\subsection{DCN and DCO$^+$ emission maps and spectra}

\begin{figure*}
\plotone{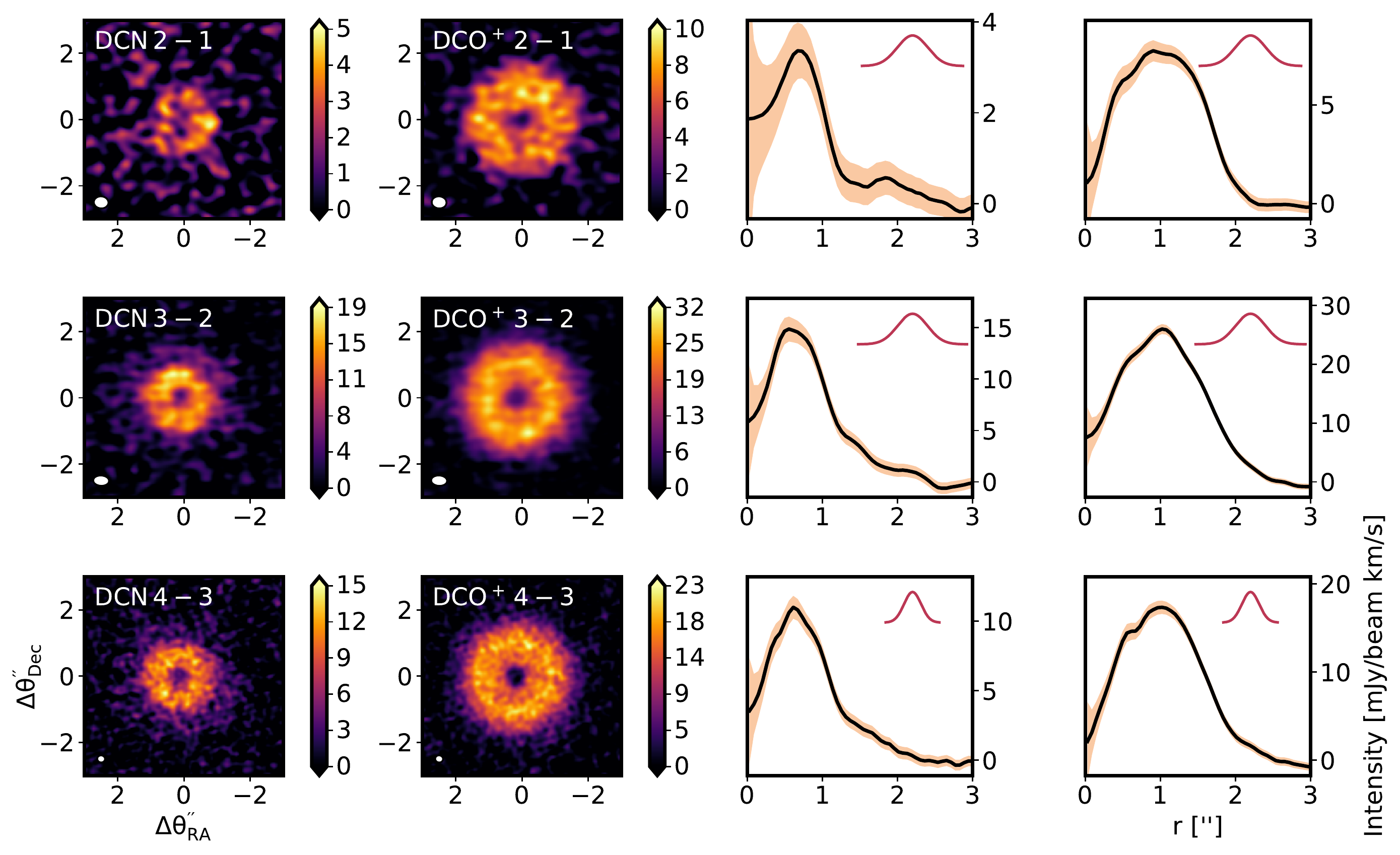}
\caption{{\it Left two panels:} Integrated emission maps of DCN and DCO$^+$ 2--1, 3--2 and 4--3 lines towards TW Hya. Beam sizes are in the bottom left corner of each panel. {\it Right two panels:} Azimuthally averaged radial intensity profiles of the same lines, with restored beam sizes plotted in each panel. Note the narrow ring + diffuse halo for DCN, and inner plateu + extended ring structure of DCO$^+$. The translucent color shows the 1$\sigma$ confidence intervals, not accounting for absolute flux uncertainties. \label{fig:mom0}}
\end{figure*}
 
Figure \ref{fig:mom0} shows integrated emission (moment-0) maps and radial profiles of the DCN and DCO$^+$ J=2--1, 3--2, and 4--3 rotational line emission toward TW Hya. The maps are constructed by integrating emission across all channels that show signal above 3$\sigma$ (Table \ref{tab:line-data}). We do not detect any molecular line hyperfine structure, i.e., it is either too weak or unresolved. The radial profiles are derived from the moment-0 maps by azimuthally integrating the inclination-corrected maps ($i=7^\circ$ and P.A.=$155^\circ$) in narrow rings. These values are slightly different from those used by \citep{Huang18}, and were selected because they provided the best visual fit to the data when overplotting a Keplerian model on top of channel maps (see Appendix \ref{app:chmaps}). Displayed uncertainties are extracted using the rms per beam in the maps and taking into account the number of independent beams in each ring. 

\begin{figure}
\epsscale{0.99}
\plotone{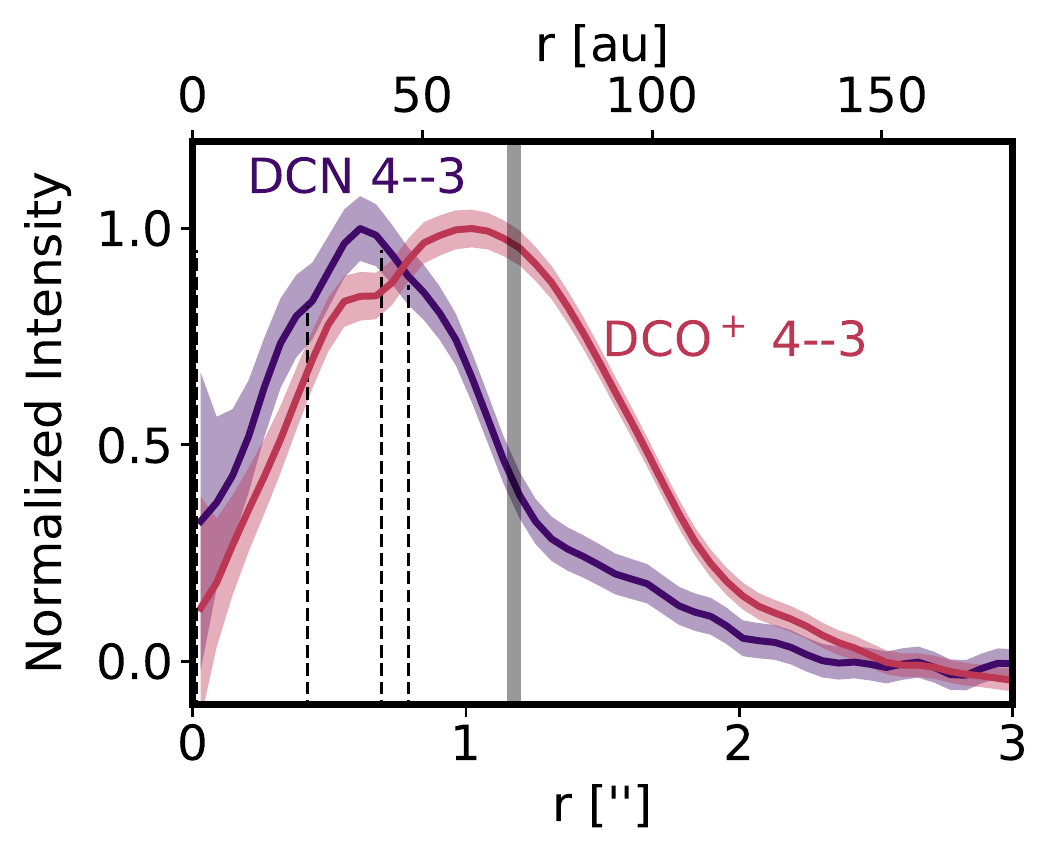}
\caption{DCN and DCO$^+$ 4--3 radial profiles normalized to the maximum flux. Note the different inner disk hole sizes. The dashed lines mark previously observed dust gaps at 25, 41 and 47~au or 0\farcs42, 0\farcs69, and 0\farcs79 \citep{Huang18}, and the broad, translucent line marks the edge of the pebble disk. \label{fig:radial}}
\end{figure}

All DCO$^+$ and DCN moment-0 maps show qualitatively similar central depressions or holes in the line emission. The central holes are not identical for DCO$^+$ and DCN, however. Figure \ref{fig:radial} shows an overlay of the higher resolution 4--3 lines, which indicate that the DCN hole has a somewhat smaller radius. On closer inspection, it seems like DCN and DCO$^+$ share a radial component that peaks at $\sim$0\farcs6. Interior to this, DCN has a shoulder at $\sim$0\farcs4. Exterior to 0\farcs6 the emission pattern clearly differs for DCN and DCO$^+$. The DCN emission decreases quickly with radius, followed by a low-intensity halo stretching out to larger radii. DCO$^+$ displays a broad second component that  encompasses the DCN halo. This second DCO$^+$ component peaks at $\sim$1\arcsec. None of these features correspond to previously noted dust rings, but we note that the DCN shoulder at $\sim$0\farcs4 or $\sim$25~au coincides with a dust gap , and a break in the $^{12}$CO radial emission profile \citep{Huang18}. Furthermore, the DCN peak nearly coincides with a second dust gap at 41~au ( $\sim$0\farcs69). The precise dust and gas properties of these gaps are unknown, and it is therefore unclear if the dust gaps are causing an increased DCN emission. One possible causal connection is that dust gaps likely present increased UV penetration, and therefore increased dissociation of CO in carbon atoms, which could fuel nitrile production. If this is the case, we would expect future observations to reveal excess emission of other nitriles at the same locations.

The emission morphologies of each molecule are consistent across the different transitions. This consistency indicates that all lines of each molecule likely originate in the same disk regions, and that observed emission substructures trace changes in column density, and not only changes in excitation conditions across the disk. We cannot rule out that some of the missing DCO$^+$ and DCN emission towards the disk center is due to continuum opacity, since \citet{Huang18} finds that the continuum $<$20~au may be, in part, optically thick. Continuum opacity cannot be the whole explanation for these central cavities, however, since other optically thin lines, including $^{13}$C$^{18}$O, present centrally peaked emission \citep{Zhang17}.  This suggests a chemical cause for the central cavities, for example the turn-off of the dominant deuterium fractionation pathway, and this is discussed in \S\ref{sec:disc}.

Table  \ref{tab:line-data} lists the integrated emission across the entire DCO$^+$ disk (out to 2\farcs5), and within a radius of 1\farcs7 or 100~au (where most DCO$^+$ and DCN emission originates). The listed uncertainties are based on the measured rms per beam in line emission-free moment-0 maps, produced from line-free channels in the relevant spw, multiplied by the square root of the number of beams within $r=$2\farcs5 and 1\farcs7, respectively. We suspect that the DCN 2--1 line emission is underestimated due to a combination of unaccounted emission from hyperfine line emission, and incomplete flux recovery of this low SNR line.

\begin{figure}
\epsscale{0.99}
\plotone{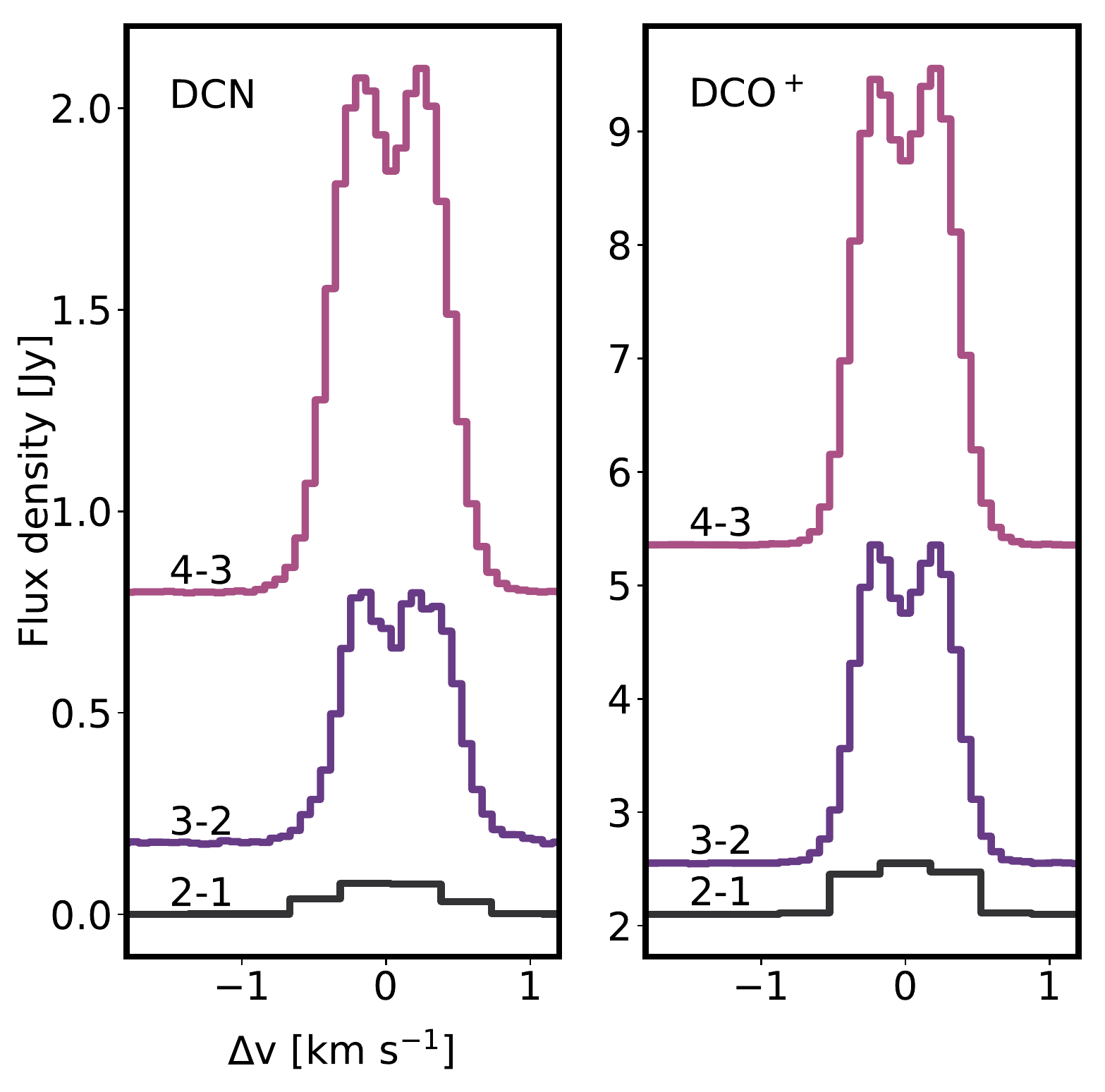}
\caption{DCN and DCO$^+$ spectra extracted using Keplerian masks. The spectra are offset for clarity. \label{fig:spec}}
\end{figure}

Figure \ref{fig:spec} shows the extracted spectra of the DCN and DCO$^+$ lines using Keplerian masks \citep{Pegues20} to enhance the SNR. The 4--3 and 3--2 lines show Keplerian profiles, while the resolution of the 2--1 lines are too poor to resolve the characteristic double-peak structure. The DCN lines are broader than the DCO$^+$ lines, which has two origins: unresolved hyperfine structure, and more emission emitting at smaller disk radii. We also inspected spectra from individual pixels and saw no evidence for substantial non-Keplerian motion or line self-absorption.
 
\subsection{Rotational diagram analysis of DCO$^+$} \label{sec:rot-dia}

\begin{figure}
\plotone{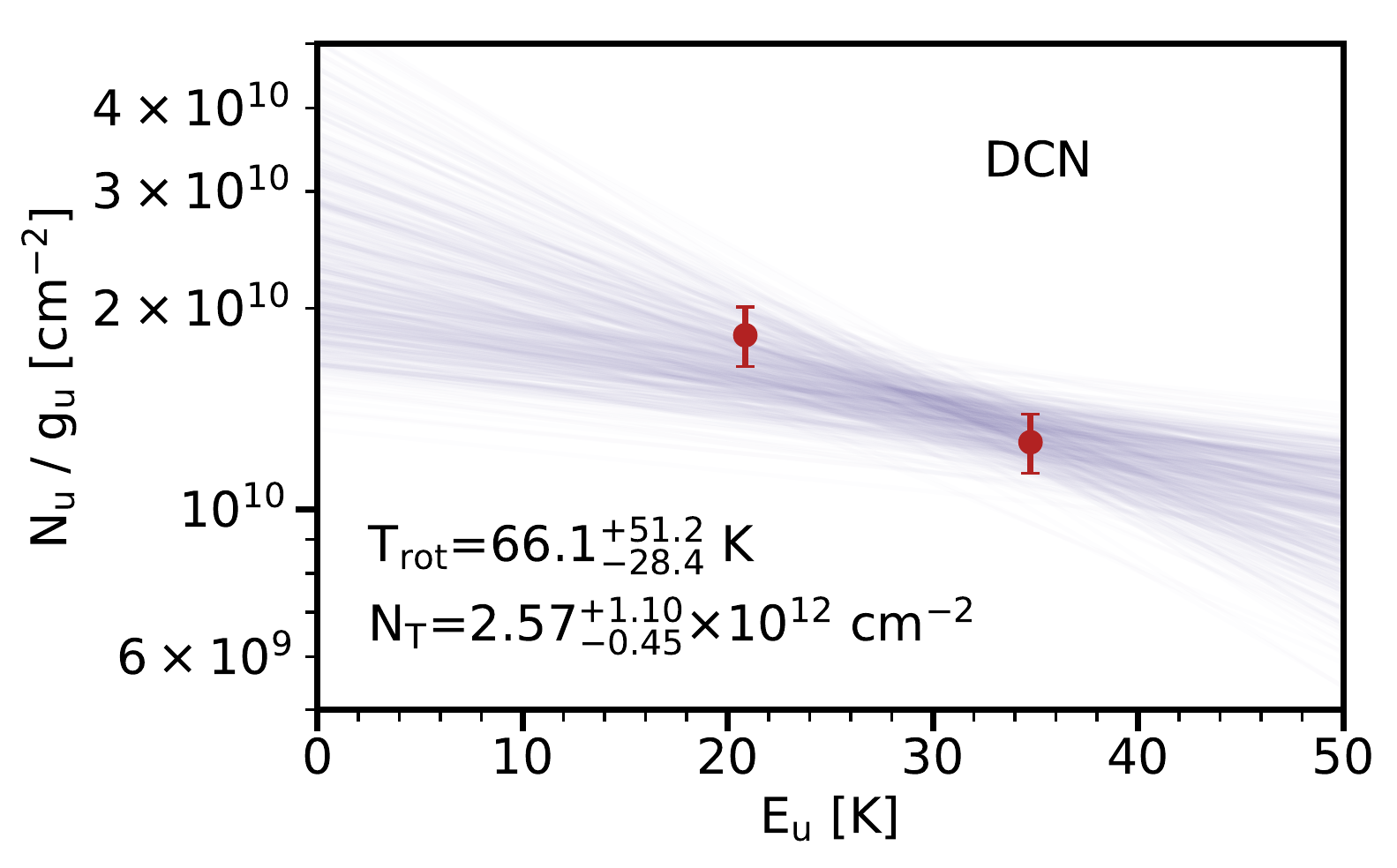}
\plotone{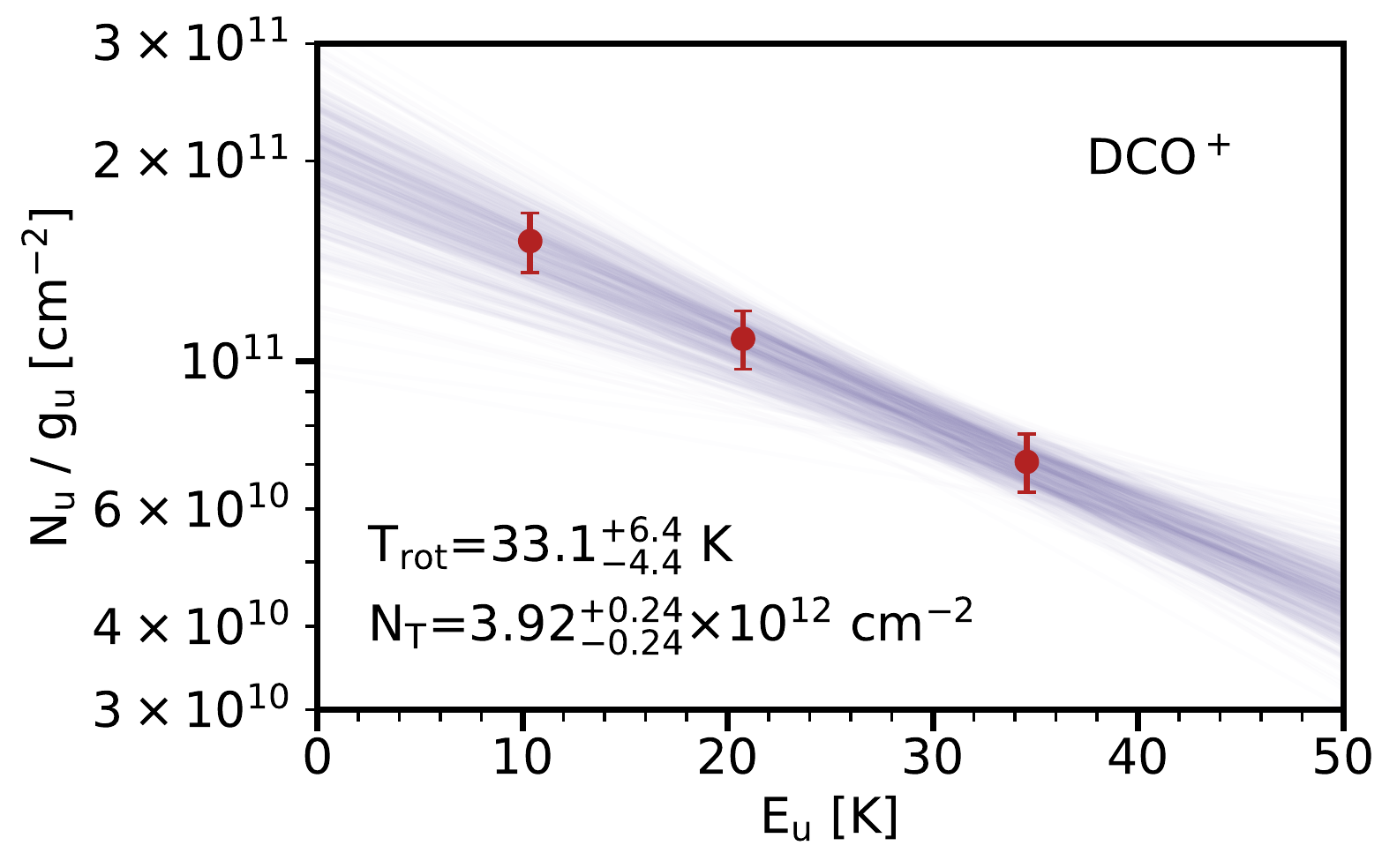}
\caption{Rotational diagram of disk-averaged DCN and DCO$^+$ lines towards TW Hya out to a disk radius of 100~au. Note that the DCN rotational diagram is very uncertain because it is based on only the 3--2 and 4--3 lines.
\label{fig:rot-dia}}
\end{figure}

We begin our characterization of DCO$^+$ and DCN using disk-averaged rotational diagrams. We use the DCN and DCO$^+$ integrated fluxes within 100~au, which encompasses the main emission features. We exclude the DCN 2--1 emission from the rotational diagram analysis because it is likely underestimated (see discussion above) -- if it is included, the derived temperature is above 100~K and the fit to the other lines is poor. The main flux uncertainty is a 10\% absolute flux calibration uncertainty, which is added in quadrature to the rms-based integrated emission errors when calculating the rotational diagram. The molecular line data was all taken from CDMS and is listed in Table \ref{tab:line-data}.  We used the following partition functions (also from CDMS) calculated at [0, 9.375, 18.75, 37.5, 75, 150, 225, 300]~K: [0, 5.769, 11.1866, 22.0293, 43.7220, 87.1365, 130.5570, 173.9803] and [0, 17.2240, 33.3906, 65.7550, 130.5095, 262.2213, 409.8604, 586.3727] for DCO$^+$ and DCN, respectively.

To calculate the rotational diagrams, we follow the MCMC  procedure outlined in \citet{Loomis18b}, using the \texttt{emcee} package \citep{Foreman-Mackey13}. 
Figure \ref{fig:rot-dia} shows rotational diagrams corresponding to random draws from the posterior
probability distributions of the excitation temperatures and column densities, with the optical depth corrected values of $N_{\rm u}/g_{\rm u}$ plotted against $E_{\rm u}$. The disk-averaged column densities of both molecules are similar: 2.6 and 3.9$\times10^{12}$ cm$^{-2}$, respectively for DCN and DCO$^+$. DCO$^+$ appears colder than DCN (33 vs 66~K), but we note that the fit to the DCN data is very uncertain since we had to remove the 2--1 line, and we cannot exclude that the actual DCN excitation temperature is $<$40~K. 

\begin{figure}[htp]
\epsscale{0.99}
\plotone{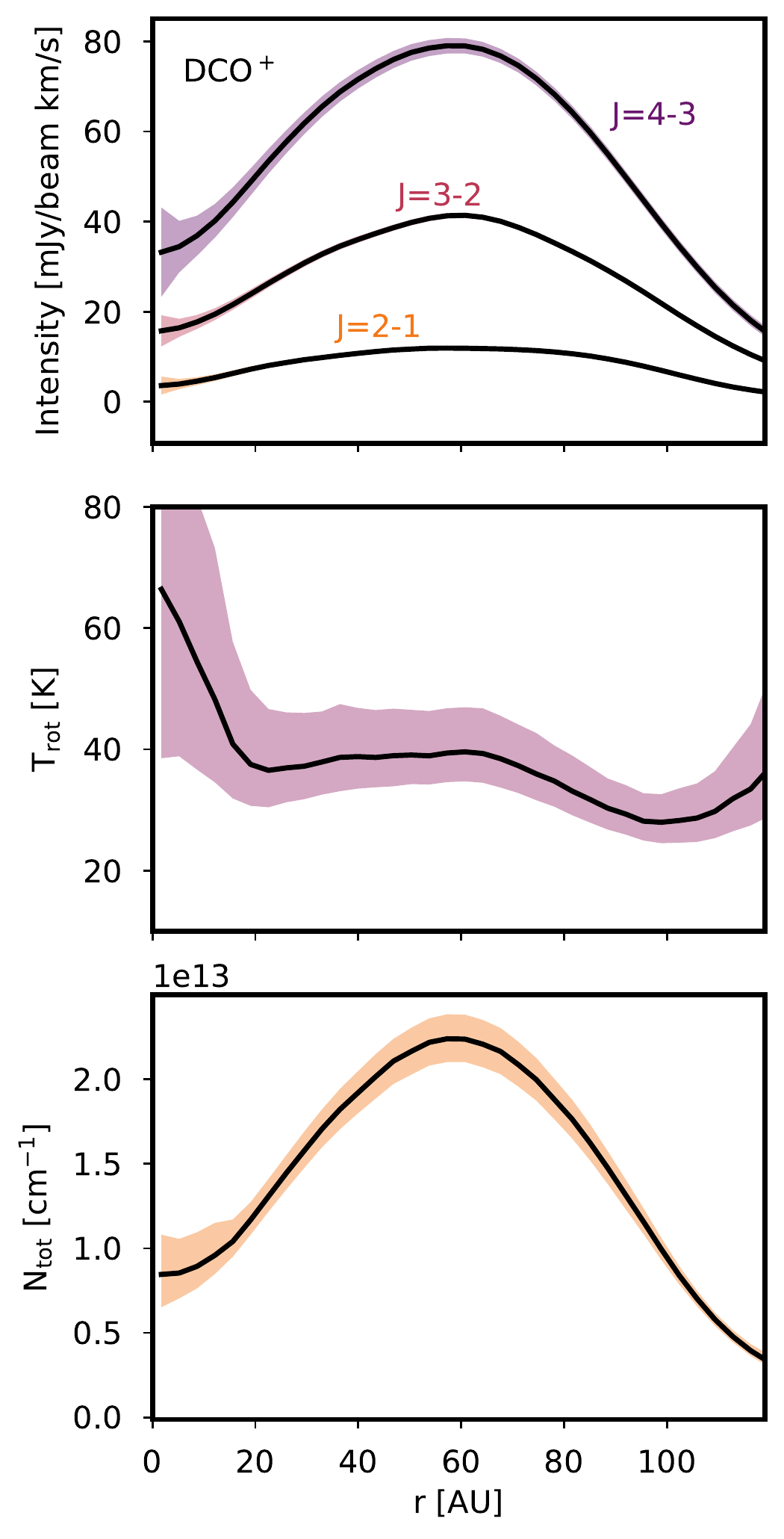}
\caption{Radially resolved rotational diagram analysis for DCO$^+$. {\it Top:} Radial emission profiles at a resolution of 0\farcs5 of DCO$^+$  lines. {\it Middle :} Derived excitation temperatures with confidence intervals corresponding to the 16th to 84th percentiles. {\it Bottom:} Calculated column density profile with confidence intervals. \label{fig:rad-rot}}
\end{figure}

We next use DCO$^+$ 4--3, 3--2, and 2--1 radial emission profiles (Fig. \ref{fig:rad-rot}, top panel) to carry out the same rotational diagram procedure in radial bins across the disk. The radial profiles were generated from moment-0 maps with a resolution of 0\farcs5, and hence appear smoother than the profiles in Fig. \ref{fig:mom0}. Similar to the disk-averaged rotational analysis, the absolute calibration uncertainty was added in quadrature to the rms-based uncertainty shown in Fig. \ref{fig:rad-rot}. The middle and bottom panels of Fig. \ref{fig:rad-rot} show the resulting radial temperature and column density profiles. DCO$^+$ appears to display a rapid decrease in excitation temperature within the inner gap, i.e. out to $\sim$25~au, but the emission levels within the gap are low and the uncertainties too high to claim a certain trend. Between 25--70~au, the best-fit DCO$^+$ excitation temperatures are 37--39.5~K, suggesting that in the bulk of the disk, DCO$^+$ originates in a layer that is more elevated than the CO snow surface, which is expected at $\sim$20--25~K. Beyond 70~au, which coincides with the outer edge of the pebble disk \citep{Andrews12,Andrews16,Huang18}, the DCO$^+$ temperature begins to drop and reaches $<$30~K at 90~au. At even larger radii there is a possible temperature inversion, but the SNR is too low to determine that this is real. We note that the lower DCO$^+$ excitation temperature in the outer disk has a disproportional impact on the disk-averaged excitation temperature due to the relatively larger emission area of the 70--120 au portion of the disk compared to the inner 70~au, which explains why the disk-averaged DCO$^+$ excitation temperature is lower than the `characteristic' excitation temperature. 

The rotational diagram results have two important caveats. First the emission surfaces of the different DCO$^+$ transitions are not necessarily the same, and this may skew the temperature upwards. We explore below whether the DCO$^+$ emission could be consistent with a colder origin. Second, the rotational diagram analysis, as well as the parametric models below assume LTE, and if some of the DCO$^+$ is originating from very elevated disk layers this assumption may not hold. While we cannot exclude non-LTE excitation, we note that it typically results in sub-thermally excited lines and therefore an under-prediction of the kinetic temperature, and should therefore not impact the main result here that DCO$^+$ appears to originate from a warmer disk layer than expected.

\section{Exploratory parametric models \label{sec:models}}

To explore what radial and vertical DCO$^+$ and DCN distributions can qualitatively explain the observations presented above, we construct a series of toy models with parametric DCN and DCO$^+$ abundance profiles. 
We first construct a simple disk density model using the common power-law prescription for the gas surface density and calculate the density using a radially dependent scale height to simulate disk flaring:

\begin{eqnarray}
\Sigma_{\rm gas} = \Sigma_{\rm gas,R=20au}\times (R/20{\rm \: au})^\gamma, \\
\rho= \frac{\Sigma}{\sqrt{2\pi}H}{\rm exp}\left(-\frac{1}{2}\left(\frac{z}{H}\right)^2\right)\\
H = H_{\rm R=20au}\times (R/20{\rm \: au})^p.
\end{eqnarray}

The surface density normalization $\Sigma_{\rm gas,R=20au}$ and power law index $\gamma$ are set to 35~g cm$^{-2}$, and $-1.3$, respectively to mimic the surface density model presented in \citet{Cleeves15}. Following \citet{Cleeves15}, we set the scale height normalization factor $H_{\rm R=20au}=2$~au, and the flaring index $p=0.3$. We adopt a simple power-law for the disk midplane temperature, using the normalization factor and power-law index derived in \citet{Zhang17}. To convert from density to number density, we adopted a mean molecular weight of 2.37, which takes into account that the hydrogen is mainly molecular.
In elevated disk layers we parameterize the gas temperature using the common prescription from \citet{Dartois03}, which takes into account direct gas heating in the lower density upper disk layers:

\begin{eqnarray}
T_{\rm mid} = 40 \times (R/10{\rm \: au})^{-0.47}\\
 T_{\rm atm} = 125 \times (R/10{\rm \: au})^{-0.47},\\
T_{R,z} = T_{\rm atm} + (T_{\rm mid}-T_{\rm atm})\times {\rm cos}(\pi z/(8H))^4 
\end{eqnarray}

\begin{figure}
\epsscale{1.2}
\plotone{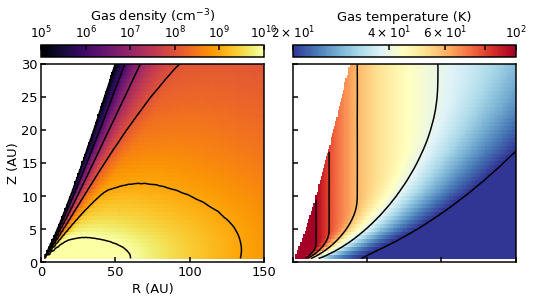}
\caption{Parametric number density and temperature distributions used to qualitatively evaluate different DCO$^+$ and DCN abundance models in the TW Hya disk.\label{fig:disk-model}}
\end{figure}

The normalization temperature of the atmosphere of 125~K follows \citet{Huang18} and is in reasonable agreement with the atmospheric temperature used by \citet{Cleeves15}. The resulting gas density and temperature structures are shown in Fig. \ref{fig:disk-model}.

\begin{deluxetable*}{llcccccc}
\tablecaption{Toy model parameters\label{tab:models}}
\tablehead{
\colhead{Model}   & \colhead{Mol.}    & \colhead{$n_{\rm R=25au}$ [$n_{\rm H}$]} &
\colhead{PL index\tablenotemark{a}}    & \colhead{$R_{\rm hole}$ [au]} &
\colhead{T boundaries [K]} & \colhead{Special constraints}}	
\startdata
A1&DCO$^+$&$1.2\times10^{-12}$&0&25&--&--\\
&DCN&$1.2\times10^{-12}$&0&25&--&--\\
A2&DCO$^+$&$8\times10^{-13}$&0.75&25&--&--\\
A3&DCN&$1.2\times10^{-12}$&0&25&$T>20$&--\\
B1a&DCO$^+$&$3.5\times10^{-12}$&0&0.1&20$<$T$<$27&--\\
B1b&DCN&$2.4\times10^{-12}$&0&0.1&20$<$T$<$30&--\\
B2&DCO$^+$&$3.5\times10^{-12}$&0&0.1&20$<$T$<$27&at $R>60$~au, T$<$27~K\\
\enddata
\tablenotetext{a}{Power-law index for abundance profile}
\end{deluxetable*}

Using this simple parametric disk structure model, we evaluate different parametric DCO$^+$ and DCN abundance models that are based on a combination of power-law prescriptions, radial cut-offs, and temperature boundaries as tabulated in Table \ref{tab:models}. In each case we simulate noise-less ALMA observations for the 2--1 and 4--3 lines using \texttt{RADMC-3D} \citep{Dullemond12}, for the radiative transfer, \texttt{vis\_sample} for the visibility sampling \citep{Loomis18a}, and then \texttt{tclean} in \texttt{CASA}. We used the DCO$^+$ molecular file from  LAMDA \citep{Schoier05}, and created our own DCN molecular file using frequencies and energy level data from CDMS \citep{Muller05}. We also included the 3--2 lines in initial model runs, but found that they did not add much to this qualitative model-data comparison due to the small difference in energy levels between the 4--3 and 3--2 lines. The simulated observations are analyzed using the same procedure as applied to our ALMA observations to enable direct comparison between observed and simulated line emission radial profiles. 

\subsection{Radial boundary models}

\begin{figure*}
\epsscale{0.9}
\plotone{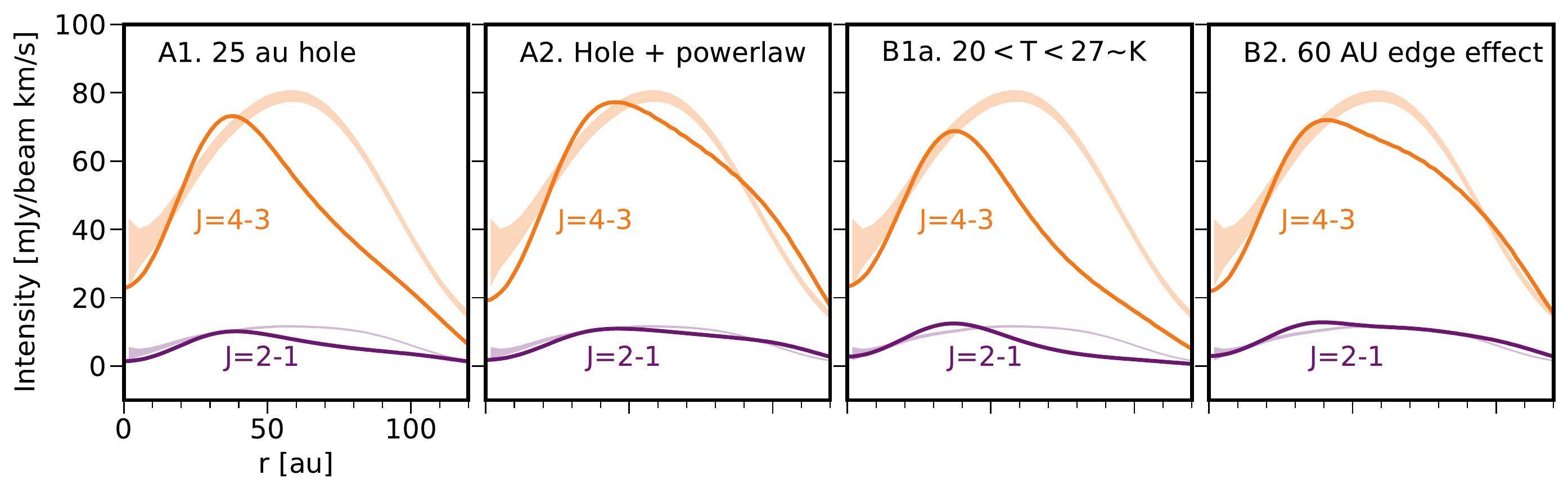}
\caption{Comparison between observed (broad bands) and modeled (thin lines) radial emission profiles of the DCO$^+$ 2--1 and 4--3 lines. The band widths of the observed profiles  approximate the observational uncertainties, but do not include a 10\% absolute calibration error. \label{fig:dcop-models}}
\end{figure*}

\begin{figure*}
\epsscale{0.7}
\plotone{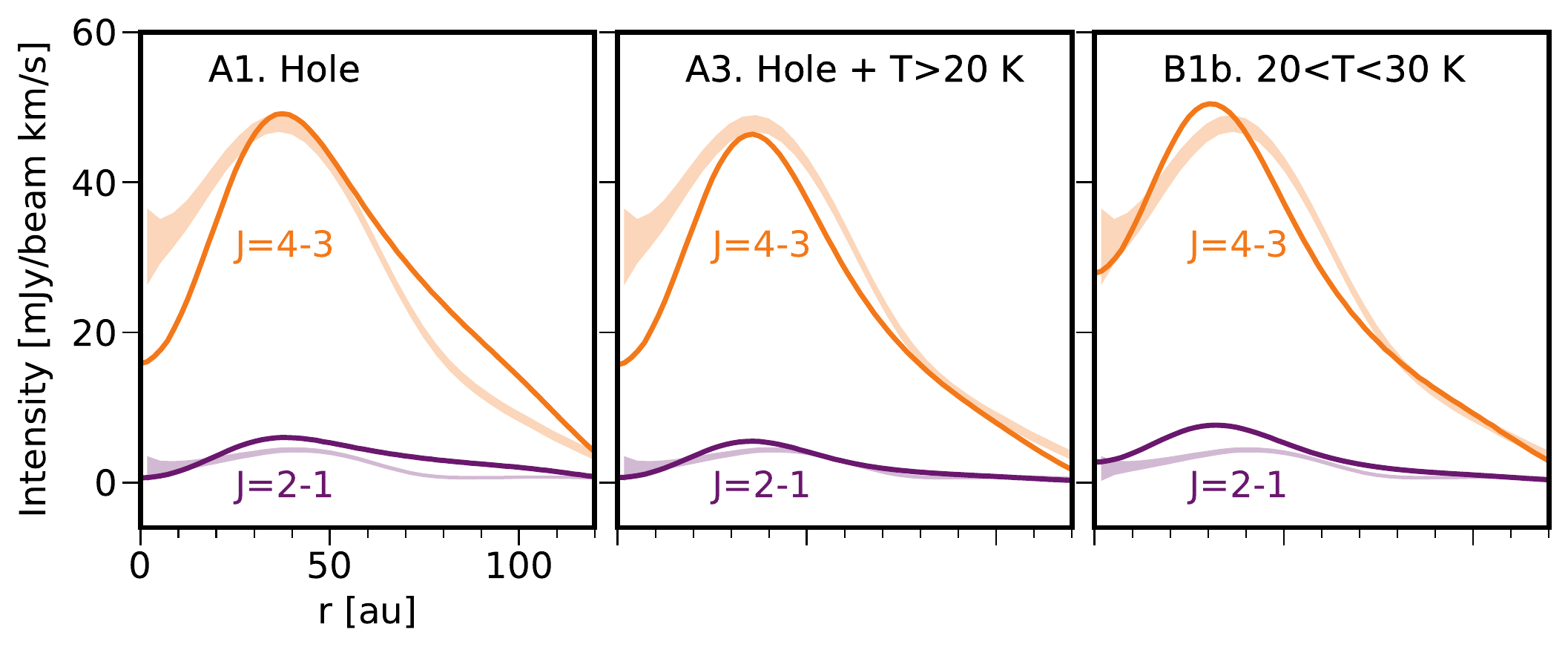}
\caption{Comparison between observed (broad bands) and modeled (thin lines) radial emission profiles of the DCN 4--3 line. The band widths of the observed profiles approximate the observational uncertainties, but do not include a 10\% absolute calibration error. Note that the observed DCN 2--1 line emission profile should be considered a lower limit for the reasons discussed in the text. \label{fig:dcn-models}}
\end{figure*}

The first set of models focuses on the inner cavity seen in both DCO$^+$ and DCN emission. The first abundance model (A1) has a constant abundance exterior to 25~au, and five orders of magnitude lower abundance in the inner hole. The abundance exterior to 25~au is estimated by eye to fit the 4--3 lines. Figures \ref{fig:dcop-models} (first panel), and \ref{fig:dcn-models} (first panel) show that this prescription does not provide a good fit to either DCO$^+$ or DCN emission, but for different reasons. In the case of DCO$^+$, the inner part of the emission profile is well fit by Model A, while the outer disk is not, indicative of that the DCO$^+$ abundance is higher in the outer than inner disk. In the case of DCN, this model produces too little emission in the hole, and too much emission at large radii, beyond 60~au.

We first address the DCO$^+$ discrepancy by exploring whether changing the abundance profile exterior to the hole from constant to an increasing power law provides a better fit (Model A2). The second panel of Fig. \ref{fig:dcop-models} shows that a DCO$^+$ abundance power law with a power law index of 0.75 does indeed provide a reasonable fit to the DCO$^+$ emission profile when imaged at this resolution. Note, however, that our higher resolution data shows a DCO$^+$ double-peak, rather than a broad single ring, which could not be explained by such a continuous power law. We also note that the model that fits the 4--3 data predicts a 2--1 emission level that is close to the one observed. This shows that the our observations do not rule out the presence of DCO$^+$ close to the disk midplane, as long as there is a substantial amount of DCO$^+$ in the warm, upper disk layers.

The DCN A1 discrepancy indicates that the DCN abundance is lower in the outer than the inner disk if DCN is emitting from all disk layers. One possible explanation is that DCN is only emitting from warmer gas, of which there is a limited amount in the outer disk. To test this, we apply a 20~K temperature boundary to the A1 model (Model A3), i.e. a constant abundance exterior to 25~au wherever the temperature is $>$20~K, and a five orders of magnitude lower abundance everywhere else. In our disk model, the midplane drops below 20~K at 44~au. The second panel of Fig. \ref{fig:dcn-models} shows that this model fits the 4--3 data quite well in the outer disk, but naturally does not fix the underabundance noted towards the disk center, which could be addressed either by making the inner cavity $\sim$5~au smaller or by implementing a smaller DCN depletion factor (not shown). The 2--1 data is always overpredicted, indicative of that the DCN is even warmer.

\subsection{Temperature boundary models}

In a second set of of models ,we apply temperature boundaries, rather than radial cut-offs, with the aim of exploring whether the observed emission profiles can be explained by DCN and DCO$^+$ temperature-dependent formation alone. The first temperature models (B1a and B1b) assume constant abundances within lower and upper temperature bounds across the disk. We explored several different boundaries, and found that the inner radial profile requires a maximum temperature cut-off at 25--30~K; 27~K provides the best fit for DCO$^+$ (B1a), and 30~K for DCN (B1b). We set the temperature minimum cut-off of 20~K, based on model predictions for CO freeze-out. The B1 model provides a good fit to the DCN 4--3 data, but over-predicts the 2--1 line emission. The 2--1 emission is also over-predicted for DCO$^+$ in the inner disk  by 20--30\%. This is qualitatively consistent with the results from the rotational diagram analysis, which indicated that a substantial amount of DCO$^+$ originates from temperatures above 30~K in the inner 70~au of the disk. 

In addition to the mismatch between the relative levels of 2--1 and 4--3 emission, the B1a model under-predicts the DCO$^+$ emission in the outer disk. We explore whether this mismatch can be explained by a second DCO$^+$ component around the dust edge, where UV photons may penetrate deeper into the disk, bringing excess cold CO into the gas-phase. We modify B1a such that exterior to 60~au (the pebble disk edge is around 70~au), DCO$^+$ is present $<$27~K, while interior to 60~au, the B1a model boundaries of 20$<$T$<$27~K still apply. This results in a more correct shape of the radial profile and may also explain the presence of a  double-peaked DCO$^+$ radial profile. We emphasize that neither the B1a/b or B2 models correctly predict the 4--3/2--1 line ratios in the inner disks for DCO$^+$ and DCN. Since the B1 and B2 temperature boundaries simulate a cold emission layer, this mismatch suggests that neither molecule is mainly present in the cold midplane.

\section{Discussion \label{sec:disc}}

\subsection{DCN and DCO$^+$ Radial and Vertical Structures}

In the inner regions of the TW Hya disk, both DCN and DCO$^+$ emission (and column densities) increase rapidly with increasing radius  starting around 20--25~au, corresponding to midplane temperatures of 27--30~K. In the absence of multi-line observations, this observation would have been in line with expectations, since if DCO$^+$ forms from reactions between the cold gas tracer H$_2$D$^+$ and CO, DCO$^+$ should be most abundant in the midplane just interior to the CO snowline \citep[e.g.][]{Mathews13,Aikawa18}. However, the inferred warm DCO$^+$ excitation temperature at 25~au shows that DCO$^+$ (and DCN) cannot be primarily emitting from the midplane. Instead the measured excitation temperature of $\sim$40~K at $25-60$~au places a substantial amount of the DCO$^+$ in an elevated disk layer.  This discovery could point to a relatively inefficient low-temperature H$_2$D$^+$ fractionation chemistry, and an efficient deuterium enrichment through reactions with, e.g., CH$_2$D$^+$. The latter is expected to proceed at higher temperatures than the H$_2$D$^+$ chemistry and may therefore be consistent with a 40~K excitation temperature \citep{Wootten87,Parise09,Favre15,Roueff15}.

This proposed scenario presents a new puzzle, however, which is that we do not observe much DCN and DCO$^+$ in the innermost midplane region. If we are observing a deuterium fractionation chemistry that is active at $\sim$40~K, we would naively expect abundant DCO$^+$ and DCN in the inner disk midplane regions that correspond to this temperature, roughly $\sim$10-20~au.  Exterior to the DCO$^+$ and DCN cavity, a possible solution to the observed emission pattern is that there are two DCO$^+$ and DCN production zones, cold and warm, respectively, which when observed towards a face-on disk masquerade as the luke-warm emission layer we observe. Indeed the A1 model in \S\ref{sec:models} shows that this is a possibility from an excitation point of view. Whether this scenario is chemically plausible is less clear, and we still need an explanation why  DCO$^+$ in TW Hya appears much warmer than in e.g. the HD~162936 disk \citep{Flaherty17}. In the inner disk the lack of  DCO$^+$ and DCN may in part be explained by continuum blocking out some fraction of the molecular emission, making the hole seem deeper than it really is. However, as discussed previously, we do not think that it is likely that continuum opacity alone is responsible for the central cavity. Instead we suggest that the lack of DCO$^+$ and DCN in the inner disk, and the elevated temperature and therefore elevated location of DCO$^+$ exterior to 25~au, together point towards a relatively inefficient DCO$^+$ and DCN production in the disk midplane at all disk radii.

One possible explanation for the lower than expected DCO$^+$ midplane abundances is that CO is depleted in the disk far beyond the CO freeze-out zone due to e.g. chemical processing and diffusion  \citep[e.g.][]{Meijerink09,Xu17,Schwarz18}. 
If CO depletion begins at 40~K instead of 25~K, this would explain why DCO$^+$ production is low both in the inner disk midplane, and close to the CO snow surface in the outer disk.
This explanation is supported by observations that show that the TW Hya disk is very CO-depleted throughout the disk molecular layers including in the inner disk \citep{Favre13,Kama16,Schwarz16,Zhang19}.  CO depletion could also diminish DCN formation in the same locations, since CO depletion from the gas would result in a depletion of the overall carbon reservoir that controls DCN and HCN production.  However, there is both observational \citep{Hily-Blant10} and theoretical (Long et al. subm.) evidence  that HCN (and by extension DCN) is not very sensitive to CO depletion, and this idea should therefore be considered highly speculative. DCN may be present at elevated disk layers simply because it mainly forms through the CH$_2$D$^+$ pathway.

If CO depletion controls where in a disk there is an active deuterium fractionation chemistry, the TW Hya results may be far from universal. CO depletion through either chemistry or diffusive flows is expected to become more severe with disk age, and TW Hya has an unusually old disk. By contrast, we may expect to find colder and more mid-plane oriented DCO$^+$ in disks that are less depleted in CO. Interestingly in the disk around Herbig Ae star HD 163296, which has been shown to be much less depleted in CO than TW Hya \citep{Zhang19}, the DCO$^+$ excitation temperature is low ($<$20~K) \citep{Flaherty17}. By contrast, most DCO$^+$ in the disk around Herbig Ae star HD 169142, appears to be warm \citep{Carney18}.  Additional observations towards a sample of young and old T Tauri and Herbig Ae disks would be key to resolve if the DCO$^+$ (and DCN) chemistry migrates to elevated disk layers over time, and if there are systematic differences between T Tauri and Herbig Ae disks.

A second possible explanation for  the inferred low levels of DCO$^+$ in the TW Hya disk midplane is that the disk midplane regions are chemically quenched due to a lack of ionization. Close to the CO snow surface, there may be too little ionizing radiation to drive a H$_2$D$^+$- or CH$_2$D$^+$-mediated chemistry, and the measured excitation temperature of DCO$^+$ may reflect the coldest disk layer where an ion-molecule mediated deuterium fractionation chemistry is efficient. TW Hya has been inferred to have a low level of ionization throughout most of the disk \citep{Cleeves15}, which supports this scenario. If this is the primary explanation for the lack of DCO$^+$ and DCN in the midplane, we would expect to see a decreasing DCO$^+$ temperature with radius, since ionization should increase in the outer, more tenuous disk regions. Indeed such a decrease is detected exterior to 60 au, but we note that there are also other possible reasons for this decrease in DCO$^+$ temperature, including release of cold CO into the gas-phase through photodesorption \citep{Oberg15}. Additional high-resolution DCO$^+$ and DCN observations towards disks with estimated ionization levels are needed to test this hypothesis.

Finally, we note that it is an open question whether we should also expect DCO$^+$ and DCN in the warm disk atmosphere in the inner disk. \citet{Favre15} predicts that DCO$^+$ should form abundantly in this disk region, while  \citet{Aikawa18} have come to a different conclusion. More theoretical work is needed to resolve this, but in the meantime we note that in the case of TW Hya, there is no evidence that the inner disk atmosphere is an important source of deuterated molecules.

The above discussion is relevant for both DCN and DCO$^+$. We now proceed with exploring reasons for observed differences between the two molecules. First, based on emission profiles and toy models, the DCN cavity is somewhat smaller  and/or less empty than the DCO$^+$ one. This suggests that there is at least one warm deuterium fractionation pathway that mainly affects DCN. There is tentative evidence for  this warmer formation channel is important for DCN throughout the disk, since the disk-averaged DCN excitation temperature appears higher than that of DCO$^+$. This, however, needs to be revisited with deeper DCN 2--1 observations.

In the outer disk of TW Hya, the DCN and DCO$^+$ radial profiles also diverge. While DCN presents a halo exterior to the pebble disk emission, DCO$^+$ has a much more substantial second emission component close to the edge of the pebble disk. Similar differences have been seen in other disks, most notably towards IM Lup and HD 163296 \citep{Huang17,Oberg15,Salinas17}. The origin of  a DCO$^+$ peak at the pebble disk edge is likely a result of increased penetration of UV radiation in the less shielded outer disk regions, which results in CO sublimation due to either a thermal inversion \citep{Cleeves16a}, or enhanced CO ice photodesorption \citep{Oberg15,Huang16,Aikawa18}. Cold, CO-rich gas constitute an  ideal environment for DCO$^+$ formation through the H$_2$D$^+$ channel. DCN clearly requires something in addition to this to form efficiently, though the presence of the DCN halo suggests that a small amount of DCN also forms under these conditions.  One possible avenue to test whether the DCO$^+$ and DCN in the outer disk originates with H$_2$D$^+$ would be to add observations of N$_2$D$^+$. N$_2$D$^+$ only forms through reactions with H$_2$D$^+$ and could therefore be used to map out where this pathway is active \citep[see e.g.][]{Pagani07,Salinas17,Aikawa18,Caselli20}. An important complication, is that N$_2$D$^+$ is only expected where there is substantial CO freeze-out and not seeing N$_2$D$^+$ can therefore not be used to rule out the H$_2$D$^+$ pathway. 

In summary, there is evidence for active deuterium fractionation chemistry in the TW Hya disk. However, much of the observed emission from DCO$^+$ and DCN appears to originate well above the CO snow surface, and some process, perhaps CO depletion or low disk midplane ionization, may be limiting the efficiency of the cold pathway in the lower disk layers and in the inner disk midplane.

\subsection{DCN and DCO$^+$ Column Densities and Abundances}

DCO$^+$ and DCN have been observed and characterized towards TW Hya in a number of earlier studies. For reference, we extracted disk average column densities of DCO$^+$ and DCN of 3.9 and 2.6$\times10^{12}$ cm$^{-2}$ respectively, and a DCO$^+$ peak column density of $\sim$7$\times10^{12}$ cm$^{-2}$. Both are substantially higher compared to values derived from single dish observations of 3 and $<$0.4$\times10^{11}$ cm$^{-2}$ for DCO$^+$ and DCN, respectively \citep{vanDishoeck03,Thi04}. This difference can likely be explained by beam dilution in the single dish observations. The large difference in DCN and DCO$^+$ column densities inferred from single dish observations is probably a beam dilution effect as well, since we find DCN to be  more compact than DCO$^+$.

DCO$^+$ and DCN have also been marginally resolved by \citet{Qi08} and \citet{Oberg12}, and these observations were used to derived radial column density profiles. \citet{Qi08} found a peak column density of $\sim$4$\times10^{12}$ cm$^{-2}$, close to our measurement. By contrast the estimates of the DCN disk averaged column density in \citet{Qi08} and \citet{Oberg12}, are an order of magnitude lower than we find here. Some of this may be explained by beam averaging, since the synthesized beam in \citet{Oberg12} was larger than the resolved DCN emitting region. The remaining difference can probably be accounted for by different disk temperature structure and DCN emitting layer assumptions.

DCO$^+$ column densities and abundances have also been estimated towards a handful of other disks and the results are remarkably similar to those we find towards TW Hya;  \citet{Teague15} and \citet{Qi15} found a DCO$^+$ column densities towards DM Tau and HD 163296 of $\sim10^{12}$ cm$^{-2}$, and \citet{Carney18} and \citet{Salinas18} found DCO$^+$ abundances with respect to hydrogen towards HD 169142 and HD 163296 of $0.9-1.5\times10^{-12}$ and $2-6\times10^{-12}$, respectively. 
The consistent DCO$^+$ column densities and abundances towards this sample of four disks is difficult to interpret, since the DCO$^+$ emitting layer appears quite different in e.g. TW Hya and HD 163296. Finally, \citep{Salinas17} also estimated the DCN abundance in the HD~163296 disk, and found $\sim$10$^{-12}$ per hydrogen nuclei, which is again consistent with TW Hya. 

\subsection{Model Comparison}

Model predictions of DCN and DCO$^+$ column density profiles, abundances, and emitting layers go back to the early 2000s. In a majority of models, DCO$^+$ column densities across disks are $\sim10^{12}$ cm$^{-2}$, in good agreement with the TW Hya findings \citep{Aikawa01,Willacy07,Aikawa18}, but as discussed below this may be a coincidence since the DCO$^+$ distributions in TW Hya and in fiducial model disks appear quite different. A notable exception is \citet{Favre15}, who predicted substantially higher column densities due to efficient warm DCO$^+$ formation. 
In contrast to our findings, DCN is predicted to be at least an order of magnitude less abundant than DCO$^+$ in most models \citep{Aikawa01,Aikawa02,Willacy07,Favre15}. The one exception is one model in \citet{Willacy07}, which  predicts similarly high DCN and DCO$^+$ column densities. This model includes efficient ice photodesorption, which both increases the overall gas-phase carbon reservoir, and desorbs some of the DCN that forms through grain surface chemistry in their model. We speculate that the difference between models and observations with regard to the relative DCO$^+$ and DCN abundances may be due to a high C/O ratio in the TW Hya disk, which would enhance both HCN production and the importance of the CH$_2$D$^+$ fractionation pathway. 

Models also predict shapes of radial profiles. In most models, DCO$^+$ displays a prominent inner hole, while DCN does not \citep{Aikawa01,Willacy07,Willacy09,Aikawa18}. This results in different DCO$^+$ and DCN radial profiles across the disk, in contrast to what is observed in both TW Hya and HD 163296, where DCN and DCO$^+$ appear to coincide at intermediate disk radii. This mismatch between models and observations suggests that the relative contributions of cold and warm deuterium fractionation pathways to DCO$^+$ and DCN remain to be fully worked out in disks.

\section{Conclusions \label{sec:conc}}

\begin{enumerate}
    \item DCO$^+$ and DCN 4--3, 3--2, and 2--1 have been observed at a spatial resolution of 0\farcs2--0\farcs4 towards the TW Hya disk. DCN presents a single narrow ring and a diffuse halo in all transitions, while DCO$^+$ presents a broader ring that breaks up into multiple components at high spatial resolution. The inner edges of the radial profiles of all DCN and DCO$^+$ transitions are similar, but not identical.
    \item Disk averaged rotational diagrams show that DCO$^+$ is present at luke-warm temperatures, just under 40~K, throughout most of the TW Hya disk, while DCN is likely warmer. The disk averaged column densities are $\sim$ 4 and 3$\times10^{12}$ cm$^{-2}$ for DCO$^+$ and DCN, respectively.
    \item Based on a series of parametric toy models, DCN emission is well fit by an inner 25~au (not completely empty) hole and a constant abundance outside of 25~au at temperatures $>$20~K. By contrast DCO$^+$ cannot be fit by any single constant abundance distribution, but requires an abundance model that takes into account the presence of a second cold reservoir of DCO$^+$ in the outer disk.
    \item DCN and DCO$^+$ production may share a formation pathway in the inner disk, where the radial profiles of the two molecules resemble one another; i.e. both molecules become abundant at elevated disk layers at $\sim$25~au. In these luke-warm layers hydrocarbon-mediated deuterium fractionation should be efficient, though we cannot exclude that the H$_2$D$^+$ pathway contributes as well. DCN presents a small shoulder interior to the main radial peak, which suggests that there is a second warmer deuterium fractionation pathway that results in DCN, but not DCO$^+$, production. In the outer disk, exterior to the pebble disk, DCO$^+$ is much more abundant than DCN and also seems to exist at lower temperatures indicative of a cold, H$_2$D$^+$-mediated deuterium chemistry. 
    \item Deuterium fractionation chemistry is generally thought of as being a low-temperature process. In the case of TW Hya deuterated molecules instead appear to mainly emit from an intermediate temperature disk layer, which suggests that either CO removal from the gas-phase or a lack of ionizing radiation has diminished the deuterium chemistry in the disk midplane. 
\end{enumerate}

Deuterium fractionation in disks is complex and multi-faceted. Multi-line observations are key to constrain excitation temperatures of abundant deuterated molecules, and further, to determine under which disk conditions they form. Ideally this should be combined with direct measurements of emission layer heights in samples of moderately inclined disks to obtain conclusive data on where and through which processes different molecules can become fractionated in deuterium. We note that TW Hya is an old, and extremely CO-depleted disk, and it will be very interesting to explore whether younger disks display a similarly distributed deuterium fractionation chemistry, or whether they enable deuterium fractionation closer to the planet-forming midplane.

\acknowledgments
This paper makes use of the following ALMA data: ADS/JAO.ALMA\#2016.1.00311.S and\\ 
ADS/JAO.ALMA\#2016.1.00440.S. ALMA is a partnership of ESO (representing its member states), NSF (USA) and NINS (Japan), together with NRC (Canada), MOST and ASIAA (Taiwan), and KASI (Republic of Korea), in cooperation with the Republic of Chile. The Joint ALMA Observatory is operated by ESO, AUI/NRAO and NAOJ.
This work is supported by NRAO. The National Radio Astronomy Observatory is a facility of the National Science
 Foundation operated under cooperative agreement by Associated Universities, Inc.
This work was supported by an award from the Simons Foundation (SCOL \# 321183, K\"O). KI\"O also gratefully acknowledges support from the David and Lucille Packard Foundation.
L.I.C. gratefully acknowledges support from the David and Lucille Packard Foundation and the Johnson \& Johnson WISTEM2D Award.
J.B.B. acknowledges support from NASA through the NASA Hubble Fellowship grant \#HST-HF2-51429.001-A awarded by the Space Telescope Science Institute, which is operated by the Association of Universities for Research in Astronomy, Incorporated, under NASA contract NAS5-26555. J.T.v.S. and M.R.H. are supported by the Dutch Astrochemistry II program of the Netherlands Organization for Scientific Research (648.000.025). J.H. acknowledges that support for this work was provided by NASA through the NASA Hubble Fellowship grant \#HST-HF2-51460.001-A awarded by the Space Telescope Science Institute, which is operated by the Association of Universities for Research in Astronomy, Inc., for NASA, under contract
NAS5-26555. C.W acknowledges financial support from the University of Leeds and from the Science and Technology Facilities Council (grant numbers ST/R000549/1 and ST/T000287/1).

\facilities{ALMA}

\software{CASA \citep{McMullin07}, Astropy \citep{astropy13,astropy18}, RADMC-3D \citep{Dullemond12}}

 \appendix
 \section{Channel Maps \label{app:chmaps}}

\begin{figure}
\epsscale{0.99}
\plotone{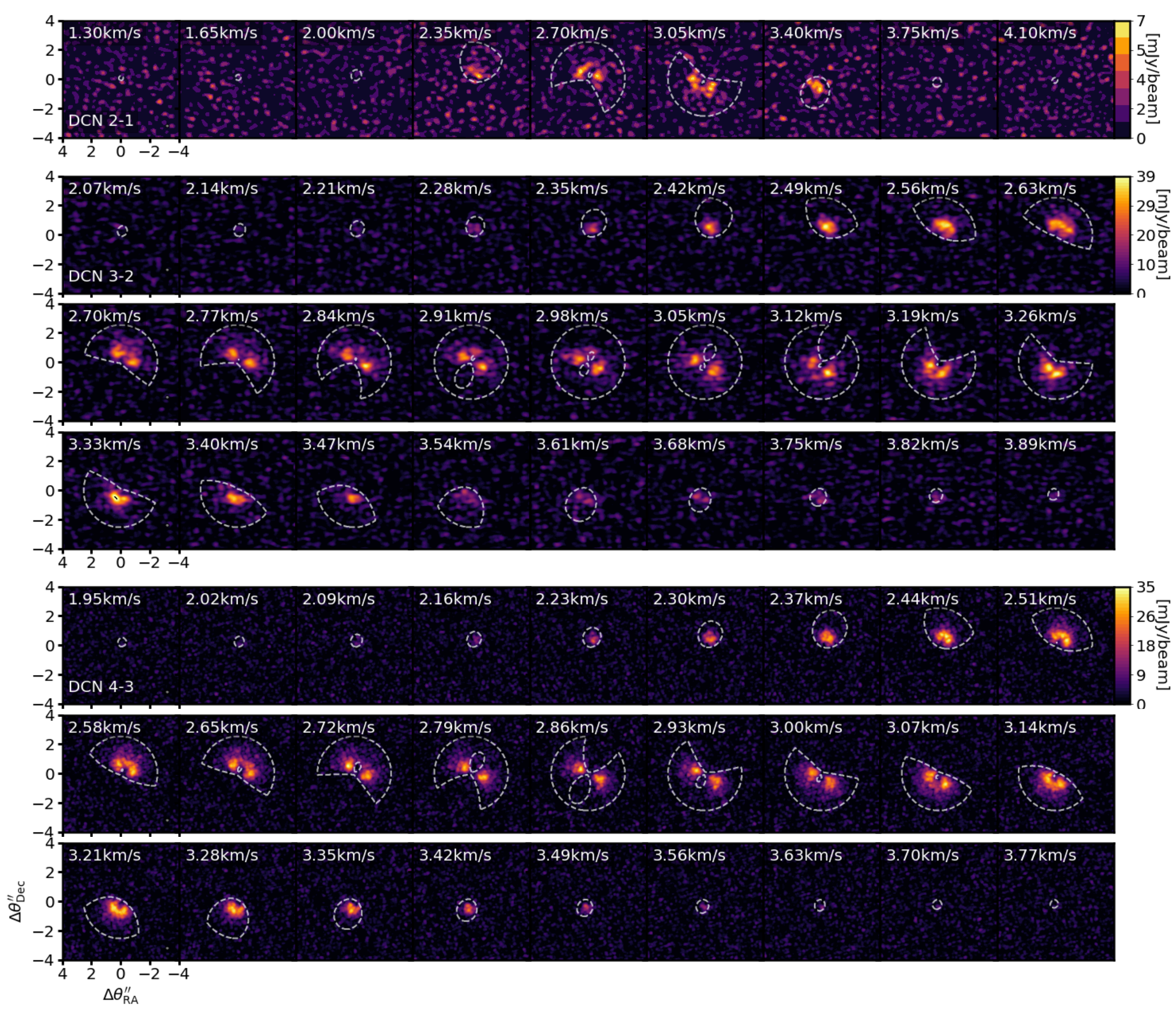}
\caption{DCN channel maps using the fiducial imaging parameters. The Keplerian mask used to extract spectra is overplotted.  \label{fig:dcn-ch}}
\end{figure}

\begin{figure}
\epsscale{0.99}
\plotone{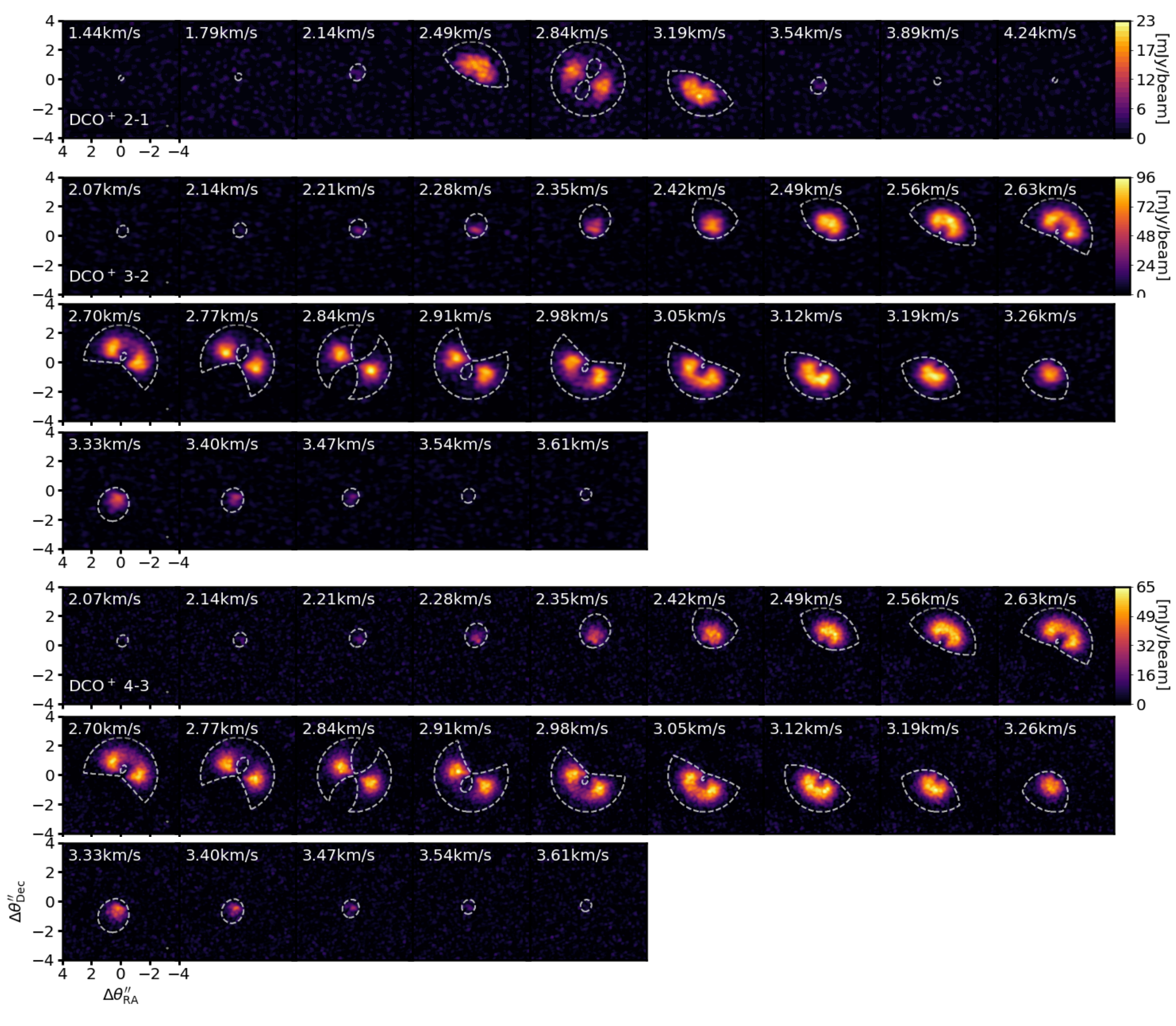}
\caption{DCO$^+$ channel maps using the fiducial imaging parameters. The Keplerian mask used to extract spectra is overplotted.  \label{fig:dco-ch}}
\end{figure}

\bibliographystyle{aasjournal}

\end{document}